\documentclass[useAMS,usenatbib]{mn2e}
\usepackage{times}
\usepackage{graphicx}
\usepackage{float}
\usepackage{amssymb}
\usepackage{amsmath}
\usepackage{color}
\usepackage{mathrsfs}
\usepackage{url}

\def\aap{A\&A}
\def\mnras{MNRAS}
\def\physrep{Phys.~Rep.}
\def\pra{Phys.~Rev.~A}
\def\nat{Nature}
\def\apj{ApJ}
\def\apjs{ApJS}

\catcode`~=11 
\newcommand{\urltilde}{\kern -.15em\lower .7ex\hbox{~}\kern .04em}
\catcode`~=13 

\title[Exact Bound-Bound Gaunt Factor]{Exact Bound-Bound Gaunt Factor Values for Quantum Levels up to n=2000}
\author[Leah K. Morabito]{\parbox{\textwidth}{Leah K. Morabito$^{1}$\thanks{E-mail: morabito@strw.leidenuniv.nl}, Gerard van Harten$^{1}$, Francisco Salgado$^{1}$, J.~B.~R. Oonk$^{2}$, H.~J.~A. R\"{o}ttgering$^{1}$, A.~G.~G.~M. Tielens$^{1}$
\\}\\
$^{1}$Leiden Observatory, P.O. Box 9513, 2300 RA, Leiden, The Netherlands\\
$^{2}$Netherlands Institute for Radio Astronomy (ASTRON), Postbus 2, 7990 AA Dwingeloo, The Netherlands\\}

\definecolor{Mygrey}{gray}{0.75}

\begin{document}

\date{}

\pagerange{\pageref{firstpage}--\pageref{lastpage}} \pubyear{2013}

\maketitle

\label{firstpage}

\begin{abstract}
Comparison of observations of radio recombination lines in the interstellar medium with theoretical models can be used to constrain electron temperature and density of the gas. An important component of the models is spontaneous transition rates between bound levels. Calculating these rates relies on accurate bound-bound oscillator strengths, which can be cast in terms of the Gaunt factor. The Gaunt factor contains terminating hypergeometric functions that cannot be calculated with sufficient accuracy for high quantum levels ($n \gtrsim 50$) by standard machine-precision methods. Methods to overcome the accuracy problem have been developed, which include asymptotic expansions and recursion relations. These methods, used in astrophysical models to calculate oscillator strengths, can introduce errors, sometimes up to as much as $\sim 8$ per cent. Detections of radio recombination lines with the new Low Frequency Array (LOFAR) has prompted an examination of theoretical models of the interstellar medium. We revisit the calculation of the Gaunt factor, employing modern arbitrary-precision computational methods to tabulate the Gaunt factor for transitions up to quantum level $n=2000$, sufficient to model low frequency Carbon radio recombination lines. The calculations provide a relative error of $\sim3\times 10^{-4}$ when compared to more detailed calculations including relativistic corrections. Our values for the Gaunt factor are provided for download in a tabular format to be used for a wide range of applications.
\end{abstract}


\section{Introduction}
Diffuse, ionized gas is one component of the interstellar medium (ISM), where ions and free electrons recombine and produce spectral lines we call recombination lines. When these transitions occur at low quantum numbers, the recombination lines appear in the optical and UV regime. 
Recombination lines occur in the radio regime when the quantum numbers involved in the transitions are above $n\gtrsim 50$, due to the decreased energy spacing of adjacent levels. Models of radio recombination lines (RRLs) therefore have to predict accurate line intensities for quantum levels above $n\gtrsim50$.

In our own Galaxy, RRLs are used to study two phases of the ISM. The \textquotedblleft classical\textquotedblright\ RRLs are associated with \textrm{H}$\,${\small\textrm{II}} regions, and are usually observed at frequencies above 1.4 GHz \citep[e.g.,][]{palmer67,roelfsema87}. These RRLs trace the warm, high density ($T\sim10^4$ K, $n_e>100$ cm$^{-3}$) medium. 
Frequencies below $\sim$1.4 GHz are associated with the cold, low-density ($T\sim100$ K, $n_e\lesssim0.05$ cm$^{-3}$) medium \citep[e.g.,][]{shaver76,payne89,ka01} and trace the diffuse component of the ISM. Diffuse Carbon RRLs with bound levels as high as $n\sim1000$ have been observed in the direction of Cassiopeia A \citep{kon80,stepkin07,asgekar13}. 
High quantum number RRLs manifest at low frequencies, and provide an important method to study physical properties such as electron temperature and density in the cold neutral medium. 
With recently completed low-frequency ($<\!\!300$ MHz, $n\!>\!280$) radio telescopes, such as the Low Frequency Array \citep[LOFAR;][]{vh13}, we will be able to study these high quantum level transitions. 
With its unprecedented sensitivity, frequency resolution and coverage and multibeaming capability enabling efficient surveys of the sky, LOFAR will revolutionise the field of low frequency RRL studies as a crucial method for studying an important phase of the interstellar medium that so far has eluded detailed studies. Not only will we be able to map a large fraction of the Galaxy, but extragalactic sources of RRLs will be accessible, providing a redshift-independent means of studying the cold, diffuse gas content of galaxies. 

The ability to calculate accurate bound-bound spontaneous transition rates in recombining ions has a wide range of applications, from predictions and measurements in the laboratory to extracting information from observations of a variety of astronomical phenomena. The transition rates depend on the atomic oscillator strengths, which are used to calculate the spontaneous and stimulated transition rates for non-relativistic electrons. 
The oscillator strength is easily described in terms either of the Gaunt factor or line strength. For low quantum number states, analytical formulae for oscillator strengths are easy to compute. However, direct calculation of higher $n$ transitions is impossible due to round-off errors and limitations on standard machine representation of numbers.

Predicting line intensities requires determining the LTE level populations of excited states, or levels, in an atom. There is a quantum level beyond which the electron will no longer be bound to the atom, and theoretical models must include all quantum levels up to this cut-off level in order to ensure correct calculations of the LTE level populations. For the low temperature, low density phase of the ISM from which we expect Carbon RRLs to originate, we have found a cut-off level of $n=2000$ is sufficient to ensure this condition and therefore provide accurate line intensities (Salgado et.~al, in prep.).

Here we revisit the calculation of the bound-bound Gaunt factor for quantum levels up to $n=2000$, using arbitrary-precision operations to compute and tabulate exact values of the Gaunt factor for easy use. 
In this paper we compare different methods of calculation, and provide a downloadable FITS table\footnote{Available via the Strasbourg astronomical Data Center, \url{http://cdsweb.u-strasbg.fr/}} for general use. In section 2 we review the analytical form for the Gaunt factor and its relation to oscillator strength. In section 3 we discuss different methods to calculate oscillator strengths. Section 4 presents our arbitrary-precision method followed in Section 5 by a comparison of various methods. Conclusions are given in section 6, and a link to the downloadable data can be found in the Supporting Information section.

\section{Oscillator Strength and the Gaunt Factor}
The Gaunt factor is used to calculate spontaneous and stimulated transition rates of electrons between quantum levels. The spontaneous transition rates, $A_{nn'}$, can be expressed directly in terms of oscillator strength,
\begin{equation}
\label{einsa}
A_{nn'} = \frac{\omega_{n'}}{\omega_n}\frac{8\pi e^2\nu^2\mu}{m_ec^3}g_{nn'}f'_{nn'},
\end{equation}
where $\omega_n$ ($\omega_{n'}$) is the statistical weight of level $n$ ($n'$), $e$ and $m_e$ are the charge and mass of an electron, $\nu$ is the frequency of the transition between level $n$ and $n'$, $c$ is the speed of light, $\mu$ is the reduced mass, and $f_{nn'}$ is the oscillator strength, which can be expressed in terms of the Gaunt factor, $f_{nn'}=g_{nn'}f'_{nn'}$. The factor $f'_{nn'}$ is calculated from the statistical weight and quantum numbers, giving a total oscillator strength of:
\begin{equation}
f'_{nn'} = \frac{2^6}{3\sqrt{3}\pi}\frac{1}{\omega_{n'}}\frac{1}{\left(\frac{1}{n'^2} - \frac{1}{n^2}\right)^3} \left\lvert \frac{1}{n^3n'^3} \right\rvert.
\end{equation}
This can be substituted into Equation~\ref{einsa}, expressing the transition rate in terms directly dependent on the Gaunt factor. Alternatively, total oscillator strength can be expressed in terms of radial dipole matrix elements, $R(nlm,n'l'm')=\langle \psi_{nlm}(\mathbf{r}) | \mathbf{r} | \psi_{n'l'm'}(\mathbf{r}) \rangle$, where $\psi$ is the hydrogenic wave function, and $\mathbf{r}$ is the electron position vector. Equation~\ref{einsa} still holds, but now the oscillator strength is defined:
\begin{equation}
\label{snn}
f_{nn'} = \frac{1}{3\omega_n}\left(\frac{1}{n^2} - \frac{1}{n'^2}\right)\frac{S_{nn'}}{e^2 a_0^2}
\end{equation}
where $S_{nn'}$ is the line strength, dependent on the radial dipole matrix elements.

The Gaunt factor for bound-bound transitions between discrete quantum levels $n$ and $n'$ have the following form \citep{mp35}, and are the same whether the transition is in absorption or emission, i.e. $g(n\rightarrow n') = g(n'\rightarrow n)$:
\begin{equation}
\label{gaunt}
g_{nn'} = \pi \sqrt{3} \left\lvert \frac{[(n-n')/(n+n')]^{2n+2n'} nn'\Delta(n,n')}{n-n'} \right\rvert .
\end{equation}
The factor $\Delta(n,n')$ is defined as:
\begin{align}
\label{delta}
\Delta(n,n') \equiv & \left[F(-n+1,-n',1,-\frac{4nn'}{(n-n')^2}\right]^2 \\
& - \left[F(-n'+1,-n,1,-\frac{4nn'}{(n-n')^2}\right]^2 \notag
\end{align}  
where $F(a,b;c;z)$ is the hypergeometric function:
\begin{equation}
\label{hyper}
F(a,b;c;z) = \sum\limits_{n=0}^{\infty}\frac{\Gamma(a+n)\Gamma(b+n)\Gamma(c)}{\Gamma(a)\Gamma(b)\Gamma(c+n)}\frac{z^n}{n!}.
\end{equation}
Expanding the series into terms provides:
\begin{equation}
\label{series}
F(a,b;c;z) = 1 + \frac{ab}{c}z + \frac{a(a+1)b(b+1)}{c(c+1)}\frac{z^2}{2} + \dots.
\end{equation}
The series will terminate, i.e. have a finite number of terms, if either $a$ or $b$ is zero or a negative integer \citep[e.g.,][]{ww63,grad07}. The magnitude of each term in the series is greater than that of the previous term, with alternating signs.

\section{Methods of Calculation}
\label{calc}
Arithmetic operations involving very large or very small numbers on computers can result in round-off errors, and the error increases with the number of operations \citep[for a more detailed discussion see e.g.,][]{nr86}. When calculating the value of Equation~\ref{delta}, the round-off errors (loss in accuracy) quickly start to add up, and once the standard floating-point limit of a machine is reached, fail to provide real, non-infinite values. This has been remarked upon by various authors \citep[e.g.,][]{gp84,delone94,fv02} and \citet{dewangan12} point out that the exact value of $n$ for which these calculations start to break down depends on the variety of algorithms and methods used. The round-off errors and limitations of the standard floating-point machine representation of numbers have driven other methods of calculation for oscillator strengths. We outline three prevalent methods in this section, and provide a summary of their accuracy and range of quantum numbers for which they are valid in Table~\ref{tab:methods}. These methods rely on numerous devices which do not involve direct calculation from the analytical formulae (given in the previous section). 

\subsection{Asymptotic Expansion}
Asymptotic expansions are powerful approximations \citep[e.g.,][]{wright35,mp35,omidvar95} that have long been used for calculations involving the Gaunt factor for quantum levels  above $n\gtrsim 50$. These expansions are still widely used in astrophysical models \citep[e.g.,][]{shaver75,sb79,kraus00,berg10}. 

A commonly cited form of the asymptotic expansion is that of \citet{mp35}, as corrected by \citet{burgess58}, which is obtained by the method of steepest descent \citep[e.g.,][]{ww63}. The first three terms are reproduced here in Equation~\ref{expansion}. This expansion is valid when the difference between the levels is large $(n'/n \ll 1)$. 
\begin{align}
\label{expansion}
g_{nn'}\simeq \, 1 & - \frac{0.1728(1+ (n'/n)^2)}{(n'(1-(n'/n)^2))^{2/3}} \\ & - \frac{0.0496(1 - \frac{4}{3}(n'/n)^2 + (n'/n)^4)}{(n'(1-(n'/n)^2))^{4/3}} + \dotsb \notag
\end{align}
This formula is easy to compute, and allows for approximations of the Gaunt factor for quantum levels that are not calculable from Equation~\ref{gaunt} using standard machine precision. These approximations have errors of $\sim0.5-8$ per cent \citep[Table 1;][]{burgess58}. \citet{omidvar95} showed that the error in the \citet{mp35} expansion can be reduced by an order of magnitude by keeping eight terms instead of five in the expansion. They additionally provide their own asymptotic expansion that has errors not in excess of $0.5$ per cent.

\subsection{Recursion Relations \label{rec}}
Another way to calculate oscillator strength is via stable recursion relations \citep[e.g.,][]{dy09,sh91,ih51}. In this method, the line strength $S_{nn'}$ in Equation~\ref{snn} is equated to the radial dipole matrix elements: 
\begin{equation}
S_{nn'} \sim  \sum\limits_{l,l'}\vert R(nl,n'l') \vert ^2.
\end{equation}
The calculation of $R$, the radial dipole matrix elements, also contains the hypergeometric function. However, the matrix elements between subsequent states can be linked through recursion relations \citep[e.g.,][]{ih51}. Given a starting point, higher quantum number states can be calculated through these relations. The total $n\rightarrow n'$ transition rate comes from summing over $l$ levels. 
\citet{sh91} use these relations to calculate values of $R(nl,n'l')$ for up to $n=500$, and provide FORTRAN code \citep[reference given in][]{sh91} to perform these calculations. \citet{dy09} also provide a FORTRAN code that makes use of recursion relations, to calculate values up to $n=1000$. Both programs were not optimized to handle calculations above these ranges, and are therefore not sufficient for use in our theoretical models of low frequency RRLs, which require calculations up to $n=2000$. 

A robust method developed by \citet[][see also Dewangan 2012]{dewangan02} exploits recursion relations in Jacobi polynomials. Standard mathematical texts transform the hypergeometric function to Jacobi polynomials, which have well known properties, including recursion relations and asymptotic expansions \citep[e.g.,][]{grad07,as72}. The recursion relations make it possible to directly calculate the Jacobi polynomials necessary, as \citet{dewangan02} demonstrated for a sampling of levels up to $n\sim1000$  using extended (quadruple) precision. The author also discusses the usefulness of the asymptotic expression of the Jacobi polynomial to examine the behaviour at large $n$, which provides results in good agreement with the analytical values for a large range of parameters. This method is discussed in far more detail in \citet{dewangan12}, and the interested reader is referred there for further details. 

\subsection{OPACITY Project}
The National Institute of Standards and Technology (NIST) Atomic Spectra Database\footnote{http://physics.nist.gov/asd} \citep{nist} is widely used by the scientific community.  This database contains, among other information, values for spontaneous transition rates that are the product of extensive calculations of the OPACITY project \citep{opacity}.
The OPACITY team extended the close-coupling method, in which wavefunctions are expanded in terms of the product of functions describing the $N$-electron states and functions describing the $(N+1)$ electron. The entire electron system is divided into an inner and an outer region, with boundary conditions set between the two regions.
An iterative process finds solutions for the outer region, and matching the boundary conditions to the inner region then provides an eigenvalue problem which can be solved to find the bound energy states. 
In the newest update to the NIST database \citep{wf07}, the team used sophisticated multi-configuration Hartree-Fock calculations that include relativistic effects to account for both fine and hyperfine structure. These values were checked against the few available experimental results, and the discrepancies between the theoretical and experimental values are less than about $2$ per cent \cite{wf07}. For a more in depth discussion of this method, we refer the reader to \citet{wf07,seaton85}, and references therein.

\setcounter{table}{0}
\begin{table*}
\caption{A Sample of Methods of Calculating Oscillator Strength.}
\label{tab:methods}
\begin{tabular}{@{}lllcl}
\hline
Method & Reference & $n$-range & Error (per cent) & Error determination \\ \hline
Asymptic Expansion & \citet{mp35} & $n\leq35$ for $n'=1,2,3$ & $\leq8$ & Comparison with selected analytical values \\
 & \citet{malik91} & $50\leq n \leq 900$ & $\dots$ & \textquotedblleft Correct\textquotedblright\ to machine precision$^1$ \\
 & \citet{omidvar95} & $\leq905$ & $0.5$ & Comparison with selected analytical values \\
Recursion Relations & \citet{sh91} & $\leq500$ & $<1$ & Evaluation of loss of significant figures \\
 & \citet{dewangan02} & $10\leq n \leq 1000$ & $\dots$ & \textquotedblleft Correct\textquotedblright\ to machine precision$^1$ \\
OPACITY Project & \citet{wf07} & $n\lesssim40$ & $2$ & Comparison with selected experimental values \\
\hline
\end{tabular} \\
$^1$ No error is reported, as transformations to equate the hypergeometric function to polynomials of other forms are used, and assumed to be exact.
\end{table*}

\section{Arbitrary-Precision Calculations}
Although the Gaunt factor (Equation~\ref{gaunt}) is of order unity, it is composed of factors that can be extremely large or small. For example, for the $n=2000$ to $n'=1999$ transition, the value of the hypergeometric functions is of the order $10^{28,000}$, while the multiplicative factor in front of it is of the order $10^{-28,000}$, yielding a value close to 1. To calculate values as large as $\sim 10^{\pm28,000}$, it is necessary to increase the precision of the calculation. {\small MATHEMATICA} \citep{mathematica} allows the user to specify the required precision of the final calculation, and uses a {\small \$MaxExtraPrecision} variable to control the precision of intermediate calculations. We requested a final precision of 10 digits, and the default value of 50 for {\small \$MaxExtraPrecision}. {\small MATHEMATICA} will keep track of the resulting precision at each intermediate step, and if the precision becomes worse than the desired precision for the final answer, {\small \$MaxExtraPrecision} will return an error. Calculating the bound-bound Gaunt factor for transitions up to $n=2000$ took approximately 5 hours in {\small MATHEMATICA} using one 2.53GHz core with 4GB of available RAM. We only performed the calculations for $n \rightarrow n'$, since the value for the inverse transition is the same. The precision in these calculations is set by the lack of relativistic corrections, and is therefore of the order $10^{-4}$. Our values therefore have a precision of $\approx 0.01$ per cent. 

\section{Comparison of Other Methods of Calculation}
\subsection{Comparison with Asymptotic Expansion}
\begin{figure*}
\begin{center}
\includegraphics[width=0.33\textwidth,clip, trim=2.5cm 0cm 0.7cm 1.1cm]{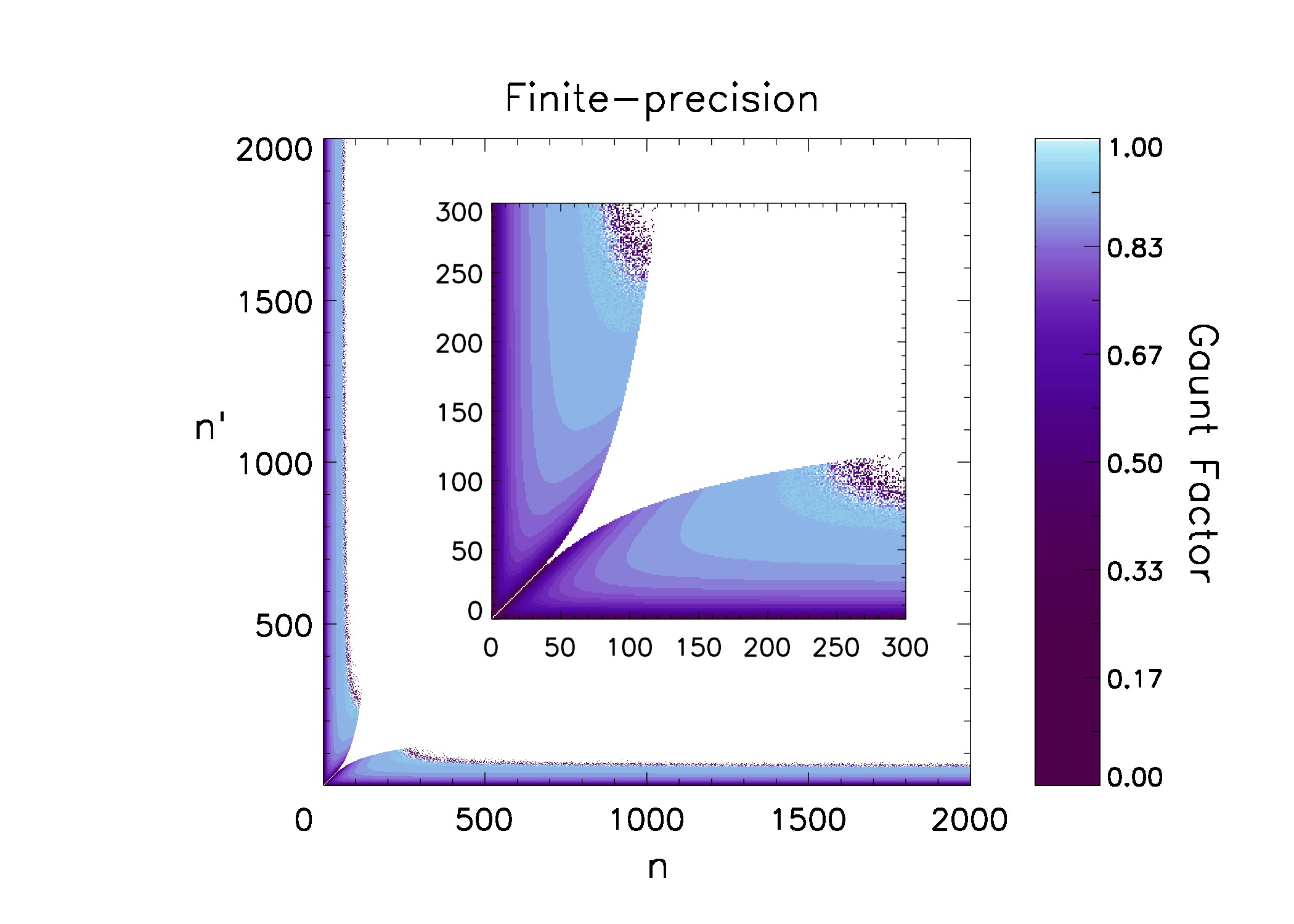}
\includegraphics[width=0.33\textwidth,clip, trim=2.5cm 0cm 0.7cm 1.1cm]{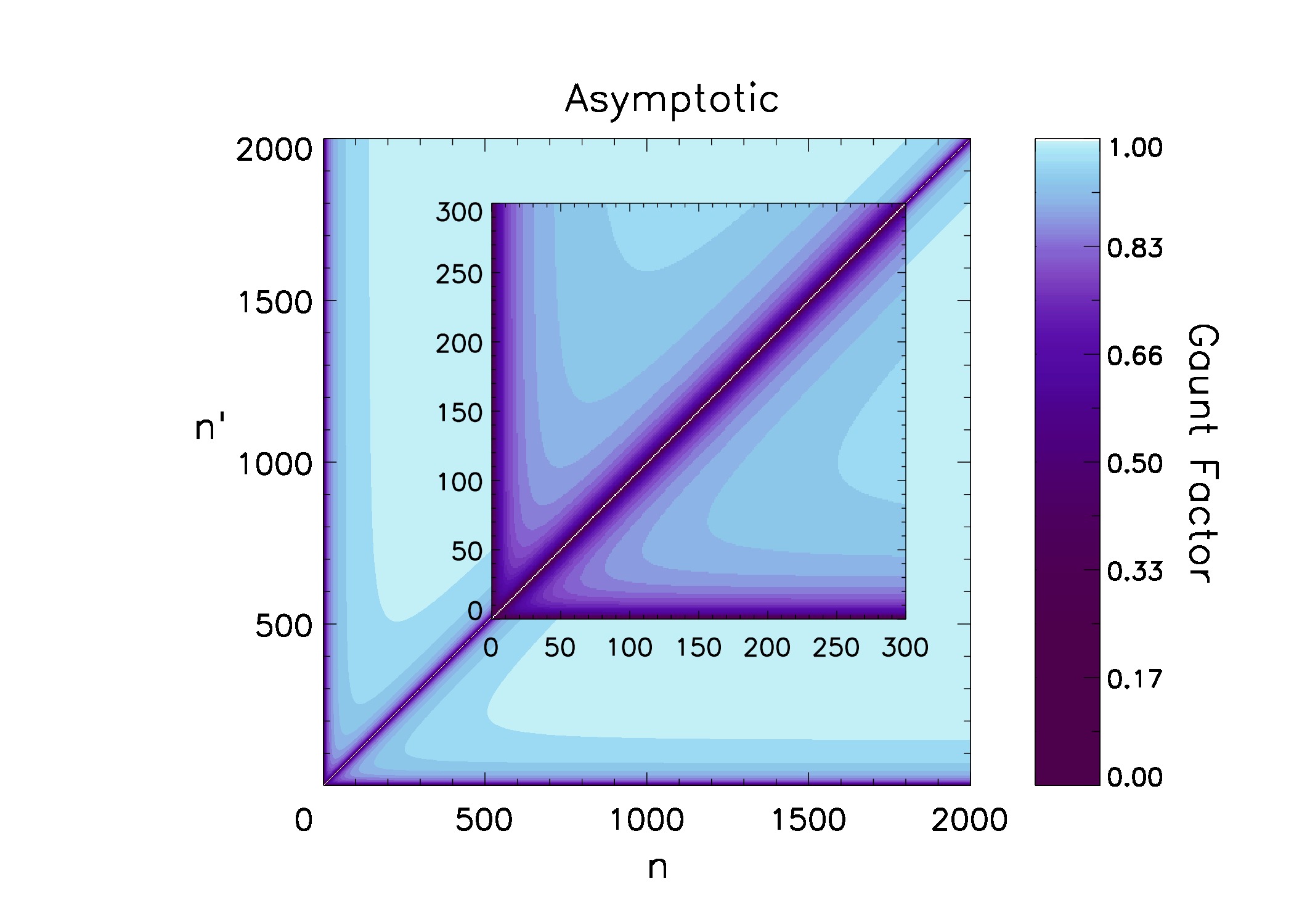}
\includegraphics[width=0.33\textwidth,clip, trim=2.5cm 0cm 0.7cm 1.1cm]{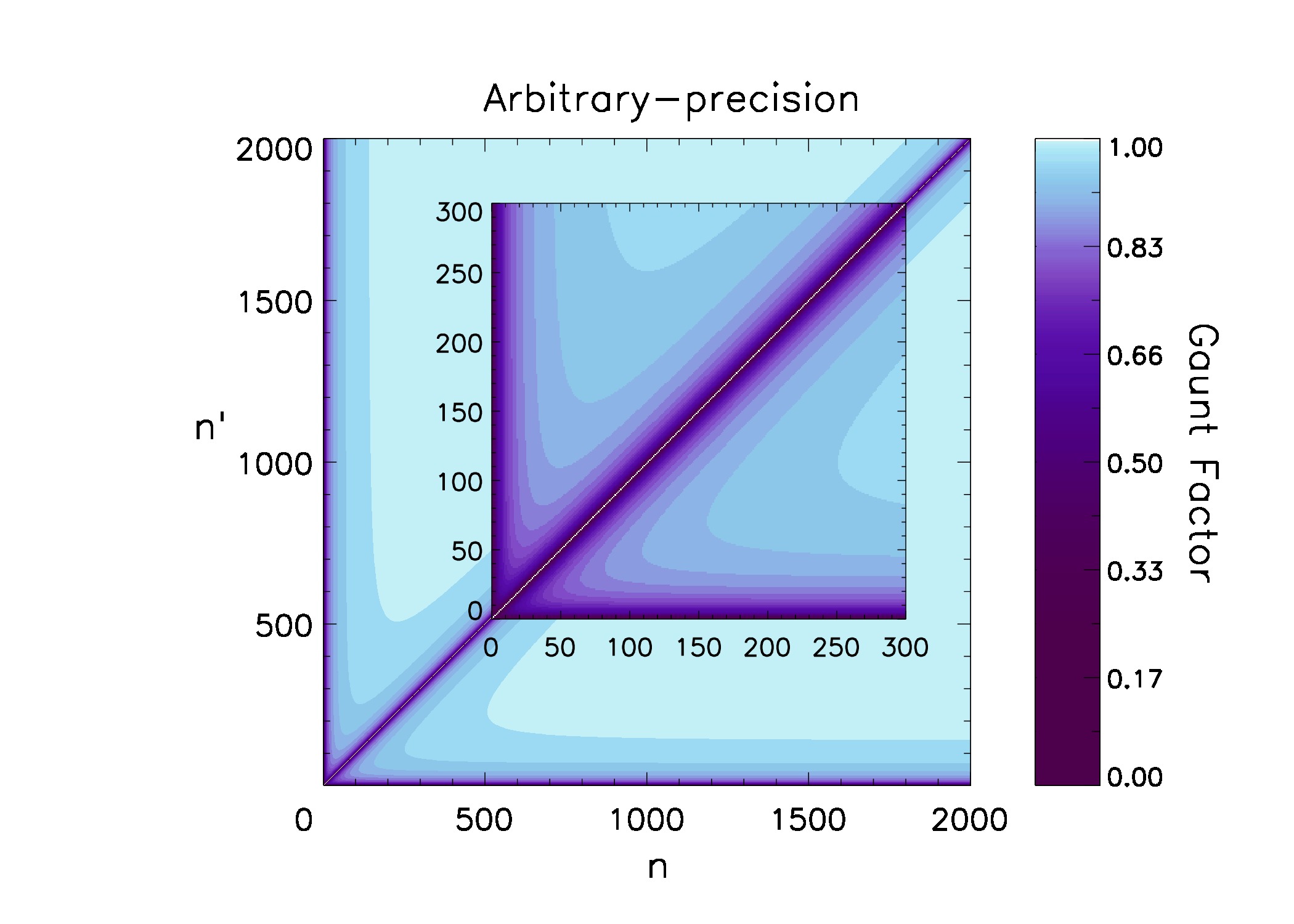}
\caption{
Values of the Gaunt factor for transitions amongst levels up to $n=2000$. From left to right: using standard-precision calculations of the analytic expression; using the \citet{mp35} asymptotic expansion; using arbitrary-precision calculations of the analytic expression in {\small MATHEMATICA}. In the left panel, the white space shows where the values of the Gaunt factor cannot be represented by standard finite-precision calculations. The middle and right panels provide values of the Gaunt factor for every transition where $n\neq n'$. The difference between the two plots is hard to distinguish, and is plotted in Fig.~\ref{facdiff}. \label{factors}}
\end{center}
\end{figure*}

\begin{figure*}
\begin{center}
\includegraphics[width=0.49\textwidth,clip, trim=2.5cm 0cm 0cm 1.7cm]{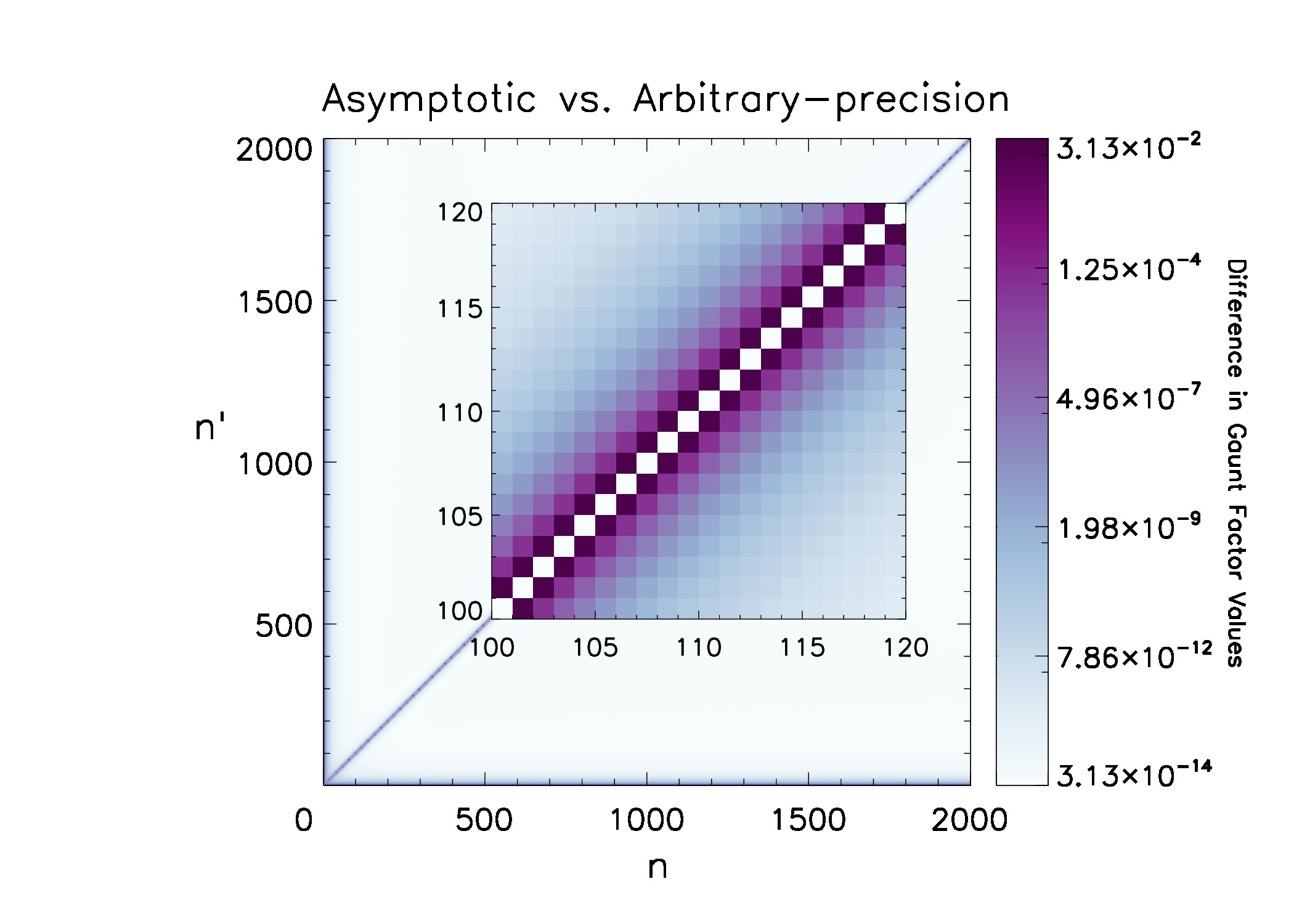}
\includegraphics[width=0.49\textwidth,clip, trim=2.5cm 0cm 0cm 1.7cm]{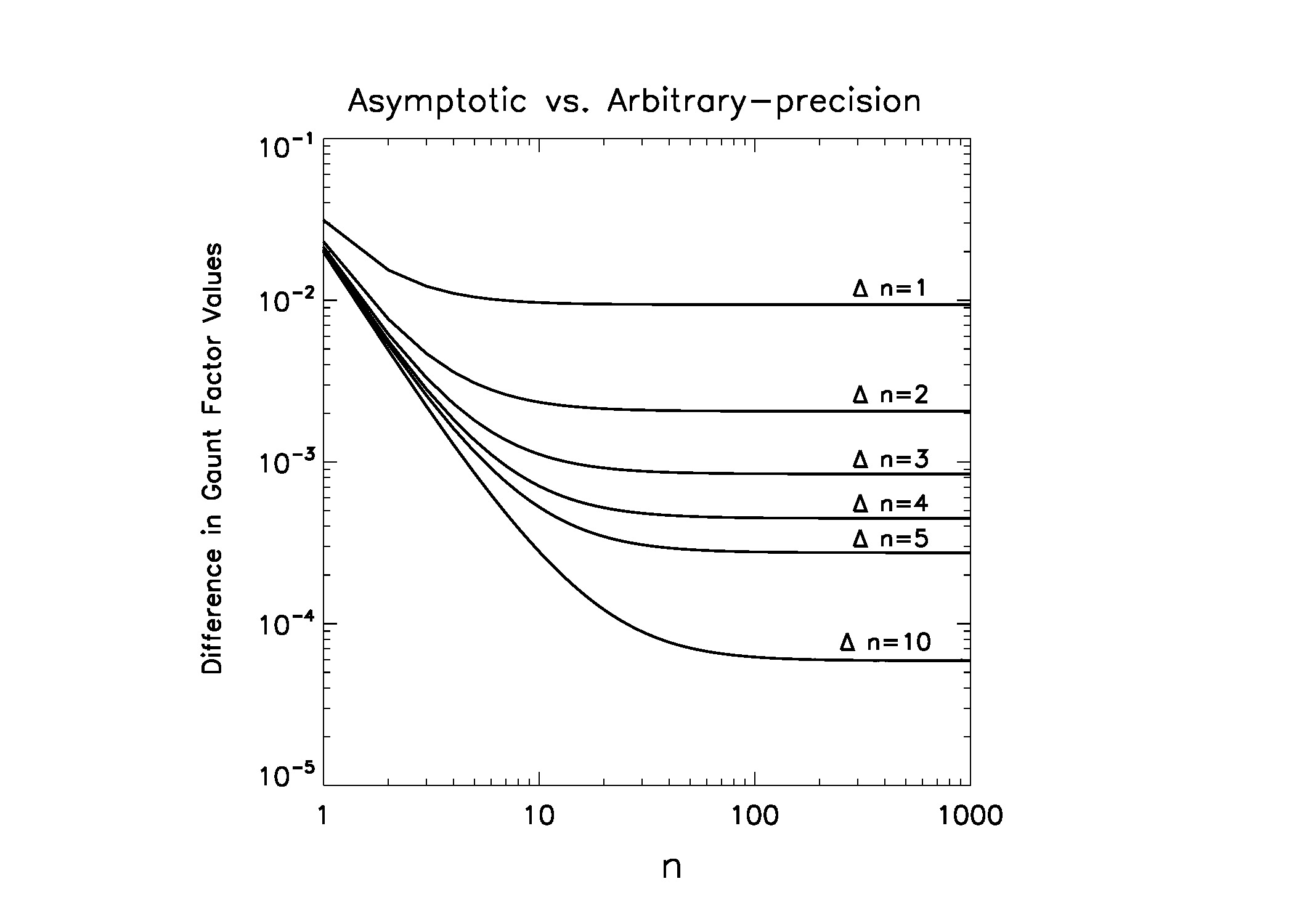}
\caption{
The absolute difference in values of the Gaunt factor as calculated by the \citet{mp35} asymptotic expansion and the analytic expression using arbitrary-precision. In the left panel, the darkest colors represent the largest differences. In the right panel, we plot the value of the difference against quantum number for several different values of $\Delta n = n-n'$. It is clear that the largest difference occurs for transitions between nearby ($n-n' \approx 1$) quantum levels. \label{facdiff} }
\end{center}
\end{figure*}

We start with a direct comparison of the Gaunt factor from the \citet{mp35} asymptotic expansion and arbitrary-precision values in Fig.~\ref{factors}. 
The top left panel shows the analytical values calculated with finite-precision. 
Starting around $n\gtrsim 50$, standard finite-precision can no longer represent the values of the hypergeometric series terms in the Gaunt factor, and therefore most of the plot is empty. 
A zoom-in of the first 300 quantum levels is shown to further clarify the behaviour on the boundary of the region where values can still be represented by standard finite-precision calculations. 
The location of this boundary is set by the amount of bits available for double-precision calculations on a machine, and to a smaller extent the algorithms and methods used. 
When round-off errors start to be large, the Gaunt factor values near the boundary fluctuate.
The top right panel demonstrates that Gaunt factor values calculated using the asymptotic expansion are at least able to fill the entire parameter space, giving real values near unity for all $\Delta n$ transitions. The bottom left panel shows the analytical values for the Gaunt factor calculated using arbitrary-precision. The plot looks remarkably similar to that of the asymptotic expansion, so we plot the difference between the arbitrary-precision and asymptotic expansion values in Fig.~\ref{facdiff}. The difference between the arbitrary-precision and asymptotic expansion values is almost always less than the accuracy in the arbitrary-precision values, but it is precisely in the region of interest, $\Delta n \sim 1$ (adjacent levels), that the difference is largest. The maximum difference is $0.03$, which means the asymptotic expansion is up to $3$ per cent too large or small compared to the analytical values.

\subsection{Relative Differences Between Methods}
\begin{figure*}
\begin{center}
\includegraphics[width=0.6\textwidth]{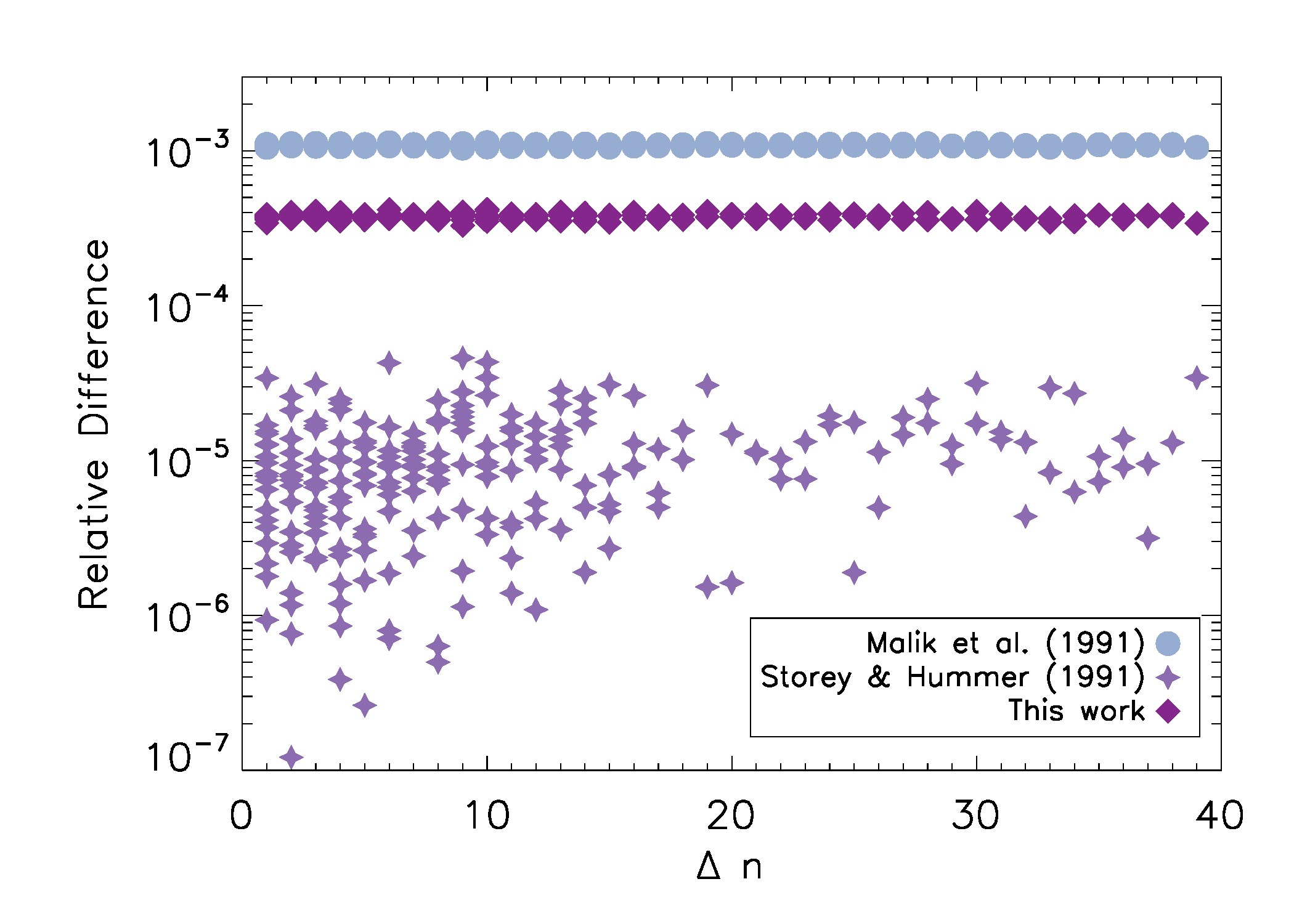}
\caption{A comparison of the relative difference in two different methods and our arbitrary-precision Gaunt factor.  The relative differences are calculated from values for spontaneous transition rates from NIST. These rates are linearly proportional to the Gaunt factor. \citet{malik91} use an asymptotic expansion, and \citet{sh91} use a recursion relation method. \label{nist}}
\end{center}
\end{figure*}

To compare with methods that calculate line strength rather than the Gaunt factor, we use the final value for spontaneous transition rates. The comparison is made using spontaneous transition rates for Hydrogen, with a reduced mass of $\mu=0.99945568$. The NIST values include relativistic corrections and are therefore more complete and precise than other methods, so we compare the relative difference between the various methods and the NIST transition rates, $|A_{n,n',\textrm{NIST}} - A_{n,n'}| / A_{n,n',\textrm{NIST}} $. We are only able to make a comparison for those levels available in NIST, and this comparison is shown in Fig.~\ref{nist}. The values from \citet{sh91} are the closest to the NIST values for these low levels, with larger scatter towards smaller changes in $n$. Although the relative difference is larger than that of \citet{sh91}, the arbitrary-precision calculations presented here only have a relative difference of only $\sim3\times10^{-4}$ from the NIST values. Therefore the arbitrary-precision values will introduce less than a tenth of per cent error into any final calculations we use them in. The scatter in the relative difference of the arbitrary-precision values of this work when compared to NIST values is of the order $10^{-5}$, indicating that our results are also stable (i.e. differences in values of $n\rightarrow n'$ have only a very small effect on the relative difference from NIST values) and predictable. 
The asymptotic expansions by \citet{malik91} and \citet{mp35} are also fairly stable, with relative differences of $1\times10^{-3}$ and $6\times10^{-3}$, respectively. 

\section{Conclusions}
We have presented arbitrary-precision calculations of the Gaunt factor for transitions up to quantum level $n=2000$, and shown that the improvement in accuracy is always at least an order of magnitude greater than the asymptotic expansions, when compared to the more complete simulations of NIST. The results of the calculations are stable and are at most only 0.03 per cent different from calculations that include relativistic corrections. The results are available for download as a FITS table. The Gaunt factor can be used for any atom to calculate spontaneous transition rates, and are therefore suitable for use by anyone working with recombination line spectra. In particular, these values fold linearly into our models of the cold neutral medium, proportionately propagating the improvement in spontaneous and stimulated transition rates. 

The data is available as a FITS table from \url{http://cdsarc.u-strasbg.fr/viz-bin/qcat?J/MNRAS/441/2855}.

\section*{Acknowledgements}
LKM acknowledges financial support from NWO Top LOFAR project, project n. 614.001.006.

\label{lastpage}

\end{document}